\documentclass[orivec]{llncs}

\usepackage{amsmath}
\usepackage{amssymb}
\usepackage{acronym}
\usepackage{graphicx}
\usepackage{xspace}
\newcommand{\req}{\stackrel{\scriptsize rank}{=}}

\newcommand{\tr}{\mathrm{tr}}
\newcommand{\ket}[1]{|#1\rangle\xspace}
\newcommand{\bra}[1]{\langle #1|\xspace}
\newcommand{\proj}[1]{\ket{#1}\bra{#1}}
\newcommand{\inn}[2]{\bra{#1}#2\rangle}

\acrodef{QT}{Quantum Theory}
\acrodef{SVD}{Singular Value Decomposition}
\acrodef{GVSM}{Generalized Vector Space Model}
\acrodef{VSM}{Vector Space Model}
\acrodef{LSI}{Latent Semantic Indexing}
\acrodef{NMF}{Non-negative Matrix Factorization}

\begin{document}

\title{Looking at Vector Space and Language Models for IR using Density Matrices}
\author{Alessandro Sordoni \and Jian-Yun Nie}
\institute{DIRO, Universit\'e de Montr\'eal \\
Montr\'eal, H3C 3J7, Qu\'ebec \\
\email{sordonia@iro.umontreal.ca},\,\email{nie@iro.umontreal.ca}}
\maketitle

\begin{abstract}
In this work, we conduct a joint analysis of both Vector Space and Language Models for IR using the mathematical framework of Quantum Theory. We shed light on how both models allocate the space of density matrices. A density matrix is shown to be a general representational tool capable of leveraging capabilities of both VSM and LM representations thus paving the way for a new generation of retrieval models. We analyze the possible implications suggested by our findings.
\end{abstract}

\section{Introduction}
Information Retrieval (IR) has nowadays become the focus of a multidisciplinary research, combining mathematics, statistics, philosophy of language and of the mind and cognitive sciences. In addition to these, it has been recently argued that IR researchers should be looking into particular concepts borrowed from physics. Particularly, it was first evoked in 2004 in Van Rijsbergen's pioneering manuscript ``The Geometry of Information Retrieval''~\cite{van_rijsbergen_geometry_2004} that Quantum Theory principles could be beneficial to IR.

Despite~\ac{QT} being an extremely successful theory in a number of fields, the idea of giving a quantum look to Information Retrieval could be at first classified as unjustified euphoria. However, the main motivation for this big leap should be found in the powerful mathematical framework embraced by the theory which offers a generalized view of probability measures defined on vector spaces. Events correspond to subspaces and generalized probability measures are parametrized by a special matrix, usually called \emph{density matrix} or \emph{density operator}.

From an IR point of view, it is extremely attractive to deal with a formalism which embraces probability and geometry, those being two amongst the pillars of modern retrieval models. Even if we believe that a unification of retrieval approaches would be out-of-reach due to the intrinsic complexity of modern models, the framework of QT could give interesting overlooks and change of perspective thus fostering the design of new models. The opening lines of Van Rijsbergen manuscript perfectly reflect this interpretation: ``It is about a way of looking, and it is about a formal language that can be used to describe the objects and processes in Information Retrieval''~\cite{van_rijsbergen_geometry_2004}.
To this end, the last chapter of Van Rijsbergen's book is mainly dedicated to a preliminary analysis of IR models and tasks by means of the language of QT. Amongst others, the author deals with coordinate level matching and pseudo-relevance feedback.

Since then, the methods that stemmed from Van Rijsbergen's initial intuition provided only limited experimental evidence about the real usefulness and effectiveness of the framework for IR tasks~\cite{piwowarski_what_2010,zhao_novel_2011,zuccon_use_2011}. Several proposed approaches took inspiration from the key notions of the theory such as superposition, interference or entanglement. In~\cite{zuccon_qprp_2010}, the authors use interference effects in order to model document dependence thus relaxing the strong assumption imposed by the probability ranking principle (PRP). An alternative solution to this problem has been proposed in~\cite{zhao_novel_2011}, in which a novel reranking approach is proposed using a probabilistic model inspired by the notion of quantum measurement. In~\cite{piwowarski_what_2010}, the authors represent documents as subspaces and queries as density matrices. However, both documents and queries are estimated through passage-retrieval like heuristics, i.e. a document is divided into passages and is associated to a subspace spanned by the vectors corresponding to document passages. Different representations for the query density matrix are tested but none of them led to good retrieval performance.

In order to give a stronger theoretical status to QT as a necessary or more general theory for IR, some authors step back into more theoretical considerations exposing potential improvements achievable over state-of-the-art models~\cite{widdows_quantum_2003,melucci_investigation_2010,melucci_deriving_2012,piwowarski_summarization_2012}. 
In~\cite{melucci_deriving_2012}, the author shows how detection theory in QT offers a generalization of the Neyman-Pearson Lemma (NPL), which is shown to be strictly linked to the PRP. Dramatic potential improvements could be obtained by switching to such more general framework. Widdows~\cite{widdows_quantum_2003} observed that the Vector Space Model (VSM) lacked a logic like the Boolean model. Through the formalism for quantum logic illustrated by Birkoff and Von Neumann~\cite{birkhoff_logic_1936}, Widdows defines a geometry of word meaning by expressing word negation based on the notion of orthogonality. Recently, the work by Melucci and Van Risjbergen~\cite{melucci_quantum_2011} and Song et al.~\cite{song_how_2010} offered a comprehensive review of QT methods for IR along with some insightful thoughts about possible reinterpretations of general IR methods (such as LSI~\cite{deerwester_indexing_1990}) from a quantum point of view. This paper shares the main purpose of the latter works.

In the ending section of his book, Van Rijsbergen calls for a reinterpretation of the Language Modeling (LM) approach for IR by means of the quantum framework. To our knowledge, such an interpretation has not been presented yet in the literature and this work can be considered as a first attempt to fill this gap. We provide a theoretical analysis of both LM and the VSM approach from a quantum point of view. In both models, documents and queries can be represented by means of density matrices. 
A density matrix is shown to be a general representational tool capable of leveraging capabilities of both VSM and LM representations thus paving the way for a new generation of retrieval models. As a conclusion, we analyze the possible implications suggested by our findings.
\section{Quantum Probability and Density Matrices\label{sec:probability_overview}}
In QT, the probabilistic space is naturally encapsulated in a complex vector space, specifically a Hilbert space, noted $\mathbb{H}^n$. 
We adopt the notation $\ket{e_1}, \ldots, \ket{e_n}$\footnote{The Dirac notation establishes that $\ket{u}$ denotes a unit norm vector in $\mathbb{H}^n$ and $\bra{u}$ its conjugate transpose.} to denote the standard basis vectors in $\mathbb{H}^n$. In QT, events are no more defined as subsets but as subspaces, more specifically as projectors onto subspaces. Given a ket $\ket{u}$, the projector $\proj{u}$ onto $\ket{u}$ is an elementary event of the quantum probability space, also called \emph{dyad}. A dyad is always a projector onto a 1-dimensional space. Generally, a unit vector $\ket{v} = \sum_i \upsilon_i \ket{u_i}$, $\upsilon_i \in \mathbb{H}$, $\sum_i |\upsilon_i|^2 = 1$, is called a \emph{superposition} of the $\ket{u_i}$ where $\ket {u_1}, \ldots, \ket{u_n}$ form an orthonormal basis for $\mathbb{H}^n$.

A density matrix $\rho$ is a symmetric positive semi-definite matrix of trace one. In QT, a density matrix defines the state of a system (a particle or an ensemble of particles) under consideration. Gleason's famous theorem~\cite{gleason_measures_1957} ensures that a density matrix is the unique way of defining quantum probability measures through the mapping $\mu_{\rho}(\proj{u}) = \tr(\rho \ket{u}\bra{u})$. The measure $\mu$ ensures that $\forall \ket{u}, \mu(\proj{u}) \ge 0$. This is because, $\mu_{\rho}(\proj{u}) = \bra{u} \rho\ket{u} \ge 0$ because $\rho$ is positive semi-definite. Moreover, if $\ket{u_1}, \ldots, \ket{u_n}$ form an orthonormal system for $\mathbb{H}^n$, the probabilities for the dyads $\ket{u_i}\bra{u_i}$ sum to one, i.e. they can be understood as disjoints events of a classical sample space. Given that $\sum_i \ket{u_i}\bra{u_i} = I_n$, the identity matrix, we have $\sum_i \tr(\rho \ket{u_i}\bra{u_i}) = \tr(\rho \sum_i\ket{u_i}\bra{u_i}) = \tr(\rho) = 1$. Therefore, for orthogonal decompositions of the vector space\footnote{\small In a more general formulation of the theory, a quantum probability measure reduces to a classical probability measure for any set $\mathcal{M} = \{M_i\}$ of positive operators $M_i$ such that $\sum_i M_i = I_n$. The set $\mathcal{M}$ is called Positive-Operator Valued Measure (POVM)~\cite{nielsen2004}. Therefore, the properties reported in this paper which apply to a complete set of mutually orthogonal projectors equally hold for a general POVM.}, a quantum probability measure $\mu$ reduces to a classical probability measure.

Any classical discrete probability distribution can be seen as a mixture over $n$ elementary points, i.e. a parameter $\vec\theta = (\theta_1, \ldots, \theta_n)$, $\theta_i \ge 0$, $\sum_i \theta_i = 1$. The density matrix is the straightforward generalization of this idea by considering a mixture over orthogonal dyads\footnote{In general, the dyads in the mixture don't need to be orthogonal. However, in this case, the coefficients $\upsilon_i$ cannot be easily interpreted as the probabilities assigned by the density matrix to each dyad.}, i.e. $\rho = \sum_{i} \upsilon_i \ket{u_i}\bra{u_i}$, $\upsilon_i \ge 0,\ \sum_i \upsilon_i = 1$. Given a density matrix $\rho$, one can find the components dyads by taking its eigendecomposition and building a dyad for each eigenvector. We note such decomposition by $\rho = R \Lambda R^\dag = \sum_{i = 1}^n \lambda_i \ket{r_i}\bra{r_i}$, where $\ket{r}_i$ are the eigenvectors and $\lambda_i$ their corresponding eigenvalues. This decomposition always exists for density matrices~\cite{nielsen2004}. Note that the vector of eigenvalues $\vec{\lambda} = (\lambda_1, \ldots, \lambda_n)$ belongs to the simplex of classical discrete distributions over $n$ points. If the distribution $\vec{\lambda}$ lies at a corner of the multinomial simplex, i.e. $\lambda_i = 1$ for some $i$, then the resulting density matrix consists of a single dyad and is called \emph{pure state}. In the other cases, the density is called \emph{mixed state}.

Conventional probability distributions can be represented by diagonal density matrices. In this case, a classical sample space of $n$ points corresponds to the set of projectors onto the standard basis $\{\proj{e_1}, \ldots, \proj{e_n}\}$. Hence, the density matrix corresponding to the multinomial parameter $\vec{\theta}$ above can be represented as a mixture, $\rho_{{\theta}} = \textrm{diag}(\vec{\theta}) = \sum_i \theta_i\ket{e_i}\bra{e_i}$. As an example, the density matrix $\rho_{{\theta}}$ below corresponds to a classical probability distribution with $n = 2$, $\sigma$ is a pure state and $\rho$ is a general quantum density, a mixed state:
\begin{equation*}
\rho_{{\theta}} = \frac{1}{2} \proj{e_a} + \frac{1}{2} \proj{e_b} = \left(\begin{matrix} 0.5 & 0 \\ 0 & 0.5 \end{matrix}\right), \;\; \sigma = \left(\begin{matrix} 0.5 & 0.5 \\ 0.5  & 0.5 \end{matrix}\right), \;\;\rho = \left(\begin{matrix} 0.5 & 0.25 \\ 0.25  & 0.5 \end{matrix}\right).
\end{equation*}

\section{Looking at Language Models}
In the Language Modeling approach to IR, each document $\mathsf{d}$ is usually assigned a unigram language model $\vec{\theta}_d = (\theta_{d1}, \ldots, \theta_{dn})$,
i.e. a categorical distribution over the vocabulary sample space $\mathcal{V}$ (of size $n$), $w \in \mathcal{V}$, $p_{{\theta}_d}(w) = \theta_{dw}$~\cite{zhai_statistical_2007}.
A query is represented as a sequence of terms $\{q_{1}, \ldots, q_m\}$, sampled i.i.d. (independent and identically distributed) from the document model.
The score for a document is obtained by computing the likelihood for the query to be generated by the corresponding document model:
\begin{equation*}
\mathcal{L}(\{q_{1}, \ldots, q_m\} | {\vec{\theta}_{d}}) = \prod_{i=1}^m p_{{\theta}_d}(q_i).
\end{equation*}
This scoring function is generally called Query Likelihood (QL). On the other hand, Kullback-Leibler (KL) divergence models can be seen as a
generalization of QL models introduced in order to facilitate the use of feedback information in Language Modeling framework~\cite{zhai_statistical_2007}.
In KL-divergence models, both documents and queries are assigned to unigram language models.
The score for a document is calculated as the negative query to document KL-divergence:
\begin{equation*}
\mathcal{KL}(\vec{\theta}_q \| \vec{\theta}_d) =  -\sum_{w} \theta_{qw} \log \frac{\theta_{qw}}{\theta_{dw}}.
\end{equation*}

\subsection{Query Likelihood View}
As presented in Section~\ref{sec:probability_overview}, conventional probability distributions can be seen as diagonal density matrices.
A straightforward quantum interpretation of the QL scoring function can be obtained by associating a diagonal density matrix to each document and consider a query as a sequence of dyads. Formally, we associate the vocabulary sample space to the orthogonal set of projectors on the standard basis, $\mathcal{E} = \{\proj{e_1}, \ldots, \proj{e_n}\}$.
The density matrix $\rho$ for a document is a mixture over $\mathcal{E}$ whose vector of weights corresponds to the parameter $\vec{\theta}_d$. Therefore, $\rho = \text{diag}(\vec{\theta}_d) = \sum_i\theta_{di} \proj{e_i}$.
 It is straightforward to show that restricted to $\mathcal{E}$, $\mu_{\rho}$ generates the same statistics as $p_{{\theta}_d}(\cdot)$, i.e. $\forall w \in \mathcal{V}$:
\begin{equation*}
\mu_{\rho}(\proj{e_w}) = \tr(\rho\proj{e_{w}}) = \sum_{i} \theta_{di}\,\tr(\proj{e_i}\proj{e_w}) = \theta_{dw} = p_{\theta_d}(w).
\end{equation*}
In the query likelihood view, the query is represented as an i.i.d. sample of word events. As word events correspond to projectors onto the standard basis, we represent a query as a sequence of i.i.d.\footnote{In quantum physics, the meaning of i.i.d. can be associated to the physical notion of measurement. If a density matrix $\rho$ represents the state of a system, an i.i.d. set of $m$ quantum events is obtained by performing a measurement on $m$ different copies of $\rho$ and by recording the outcomes.} quantum events belonging to $\mathcal{E}$, $\{\proj{e_{q_1}}, \ldots, \proj{e_{q_m}}\}$. Therefore, the score for a document is computed by the following product:
\begin{equation}
\mathcal{L}(\{\proj{e_{q_1}}, \ldots, \proj{e_{q_m}}\} | \rho) = \prod_{i=1}^m \mu_{\rho}(\proj{e_{q_i}}) =  \prod_{i=1}^m p_{{\theta}_d}(q_i),
\end{equation}
which indeed corresponds to the classical QL scoring function. However, we shall stress on an important point about the equation above. If the projectors included in the query sequence are mutually orthogonal (as above), the calculation above behaves as a proper classical likelihood, i.e. the sum of the likelihoods of all possible samples of length $m$ is one. On the contrary, the product cannot be considered as a classical likelihood in general because quantum probabilities for arbitrary events does not need to sum to one. Further considerations on these issues will be made in Section 6.

\subsection{Divergence View}
The KL scoring function computes a divergence between a query language model $\vec{\theta}_q$ and document language model $\vec{\theta}_d$.
In QT, the KL-divergence is a special case of a more general divergence function acting on density matrices called Von-Neumann (VN) Divergence. Note $\rho = \sum_i \lambda_i \proj{r_i}$, and $\sigma = \sum_i \zeta_i \proj{s_i}$ the eigendecompositions of two arbitrary density matrices. In the following, the log function applied to a matrix refers to the matrix logarithm, i.e.  the natural logarithm applied to the matrix eigenvalues, $\log \rho = \sum_i \log \lambda_i \proj{r_i}$. The VN divergence writes as:
\begin{equation*}
\mathcal{VN}(\rho \| \sigma) = \tr(\rho(\log\rho - \log\sigma)) = \sum_{i} \lambda_{i}\log\lambda_{i} - \sum_{i, j} \lambda_{i} \log\zeta_{j} |\inn{r_i}{s_j}|^2.
\end{equation*}
This divergence quantifies the difference in the eigenvalues as well as in the eigenvectors of the two density matrices~\cite{tsuda_matrix_2006}.

In order to see how the classical KL retrieval framework is recovered, we assign a density matrix to the query very similarly to what has been done for a document. Precisely, $\rho_{q}$ and $\rho_{d}$ are diagonal density matrices such that $\rho_{q} = \sum_{i} \theta_{qi} \ket{e_i} \bra{e_i}$ and $\rho_{d} = \sum_{i} \theta_{di} \ket{e_i} \bra{e_i}$. As $\rho_q$ ($\rho_d$) is diagonal in the standard basis, its eigenvalues correspond to $\vec{\theta}_q$ ($\vec{\theta}_d$), thus:
\begin{equation}
\mathcal{VN}(\rho_q \| \rho_d) = \sum_{i} \theta_{qi}\log\theta_{qi} - \sum_{i,j} \theta_{qi} \log\theta_{dj} |\inn{e_i}{e_j}|^2 =  \sum_{i} \theta_{qi}\log \frac{\theta_{qi}}{\theta_{di}},
\end{equation}
which corresponds to the KL divergence. As conventional probability distributions correspond to diagonal density matrices, their eigensystem is fixed to be the identity matrix. Intuitively, KL divergence captures the dissimilarities in the way they distribute the probability mass on that eigensystem, i.e. by their eigenvalues.


\section{Looking at the Vector Space Model\label{sec:vsm}}
In this section, we are attempting to look at the VSM~\cite{salton_term-weighting_1988} in a new way.  In its original formulation, no probabilistic interpretation could be given because of the lack of an explicit link between vector spaces and probability theory~\cite{wong_modeling_1995}. In the model, documents and queries are represented in the non-negative part of the vector space $\mathbb{R}_+^{n}$, where $n$ is the number of terms in the collection vocabulary. In VSM, each term corresponds to a standard basis vector. The location of each object in the term space is defined by term weights (i.e. \emph{tf}, \emph{idf}, \emph{tf-idf}) on each dimension. Similarity between documents and queries are computed through a vector similarity score $\vec{q}^\top \vec{d}$, where $\vec{q}, \vec{d}$ are the vector representations of the query and the document. In~\cite{salton_term-weighting_1988}, the authors show that normalizing document vectors is important to reduce bias introduced by variance on document lengths. By normalizing both document vector and query vector, the similarity score reduces to the cosine similarity between the two vectors, which is an effective similarity measure in the model~\cite{zobel_exploring_1998}. From now on, we consider $\ket{q}, \ket{d} \in \mathbb{R}_+^n$, the normalized ($\|\cdot\|_2$) query vectors. Documents can thus be safely ranked by decreasing cosine $\inn{q}{d} \in [0, 1]$, which cannot be negative because the ambient space is $\mathbb{R}_+^n$.\footnote{In this paper, we do not explicitly take into account situations in which the vectors could contain negative entries. For example, this could easily happen after the application of Rocchio's algorithm~\cite{rocchio_relevance_1971} in feedback situations or by reducing the dimensionality of the vector space by LSI~\cite{deerwester_indexing_1990}. Besides the historically encountered difficulties in the interpretation of such negative entries~\cite{hofmann_unsupervised_2001}, in these particular cases, the rank equivalence situations discussed here could not hold. However, we argue that ignoring these situations causes no harm to the generality of our conclusions on the need of an enlarged representation space.}

\subsection{Query Likelihood View}
In this interpretation of the VSM, each document is associated to a probabilistic ``model'' in the same spirit of the Language Modeling approach. We define a density matrix $\rho$ for the document as $\rho_d = \proj{d}$,
which is a pure state, i.e. its mixture weights are concentrated onto the projector $\proj{d}$. Note that this density matrix does not have a statistical meaning. It has been determined by merely normalizing heuristic weighing schemes and it cannot be related to a statistical estimators such as Maximum Likelihood (MLE).

A query can be represented as the quantum event corresponding to the subspace spanned by $\ket{q}$. This subspace naturally corresponds to the dyad $\ket{q}\bra{q}$. Hence, a query can be seen as the sequence of quantum events of length one $\{\ket{q}\bra{q}\}$. In this setting, its likelihood given the document model is calculated by:
\begin{equation}
\label{eq:vsm:proba}
\mathcal{L}(\{\ket{q}\bra{q}\} | \rho_d) = \mu_{\rho_d}(\ket{q}\bra{q}) = \mathrm{tr}(\rho_d\proj{q}) = \mathrm{tr}(\inn{q}{d}\inn{d}{q}) = |\inn{q}{d}|^2,
\end{equation}
The above calculation shows that the quantum ``likelihood'' assigned to the event $\proj{q}$ by the density $\rho_d$ is the square of the cosine similarity between the query and the document. When restricted to the non-negative domain, the square function is a monotonic, increasing transformation. This means that $\mu_{\rho_d}(\proj{q}) \req \inn{q}{d}$,
i.e. the two formulations lead to the same document ranking.


\subsection{Divergence View}
According to the original VSM, queries and documents should share the same representation and the scoring function should be a distance measure between these representations. In the previous formalization, this initial paradigm seems apparently lost. The following alternative quantum interpretation of the VSM is perhaps closer to the original vision of the model. We associate a density matrix both to the document and to the query. Specifically, those density matrices would be pure states, projectors onto the corresponding vectors, i.e. $\rho_d = \proj{d}$, $\rho_{q} = \proj{q}$. It turns out that computing the Fidelity measure~\cite{nielsen2004} between density matrices produces a ranking function equivalent to cosine similarity:
\begin{equation}
\mathcal{F}(\rho_q, \rho_d) = \tr(\sqrt{\sqrt{\rho_q}\rho_d\sqrt{\rho_q}}) = \tr(\sqrt{\ket{q}\inn{q}{d}\inn{d}{q}\bra{q}}) = |\inn{q}{d}| \tr(\rho_q) = |\inn{q}{d}|,
\end{equation}
obtained by noting that $\rho_q$ is a projector thus $\sqrt{\rho_q} = \rho_q$, and $\tr(\rho_q) = 1$. As $\ket{q}, \ket{d} \in \mathbb{R}^n_+$, ranking by Fidelity measure is equivalent to ranking by cosine similarity, thus $\mathcal{F}(\rho_q, \rho_d) \req \inn{q}{d}$.

\section{A joint analysis}
In this section, we will try to summarize the commonalities and the differences arising from the quantum formalizations of the two models given in the preceding sections. The following analysis is succinctly reported in Table 1. As a starting point, we shall note that the ambient space for both models is the Hilbert space $\mathbb{H}^n$, where $n$ is the size of the collection vocabulary. Each standard basis vector $\mathcal{E} = \{\ket{e_1}, \ldots, \ket{e_n}\}$ is associated to a word event. Therefore, the vocabulary sample space corresponds to the set of projectors onto the standard basis vectors $\{\ket{e_i}\bra{e_i}\}_{i=1}^n$.

\subsection{Query Likelihood View}
In query likelihood interpretations, the query is represented as a sequence of i.i.d. dyads. In the VSM, the sequence contains one dyad corresponding to the projector onto the query vector $\{\ket{q}\bra{q}\}$. On the contrary, in the LM approach the sequence contains a dyad for each classical word event, i.e. $\{\ket{e_{q_1}}\bra{e_{q_i}}\}_{i=1}^m$.

Besides the number of dyads included in the sequence, a major difference distinguishes the two formalizations. Contrary to probabilistic retrieval models such as LM, a query is not considered as a sequence of independent classical word events but as a single event and a particular kind thereof. The query event is a \emph{superposition} of word events. This can be seen because the vector $\ket q$ can be expressed, up to normalization, as $\ket{q} = \sum_w f(w)\, \ket{e_w}$ where $f(w)$ is the weight for term $w$ in the query vector. This kind of event neither can expressed using set theoretic operations nor it has a clear classical probabilistic interpretation: it does not belong to $\mathcal{E}$ thus it can only be justified in the quantum probabilistic space. Arguing further, we would say that, in the case of VSM, term weighting methods aim at estimating the ``best'' query event, i.e. the event which is the most representative for the information need of the user. Intuitively, if a single choice would be given to us on what to observe, we would rather be observing in the ``direction'' of important words in the query.
\begin{table}[t]
\caption{\label{table:succ}Summary of the representations for documents and queries and the scoring functions of the two studied methods.}
\def\arraystretch{1.1}
\setlength{\tabcolsep}{8pt}
\centering
\small
\begin{tabular*}{\textwidth}{llll}
\hline\noalign{\smallskip}
\multicolumn{4}{c}{\textbf{Query Likelihood View}} \\
\noalign{\smallskip}
\hline
\hline
& Query & Document & Scoring \\
\hline
\noalign{\smallskip}
VSM & $\{\proj{q}\}$ & $\rho_d = \proj{d}$ & $\mu_{\rho}(\proj{q})$ \\
LM & $\{\ket{e_{q_1}}\bra{e_{q_1}}, \ldots, \ket{e_{q_m}}\bra{e_{q_m}}\}$ & $\rho_d = \sum_w \theta_{dw}\ket{e_w}\bra{e_w}$ & $\prod_{i}\mu_{\rho_q}(\ket{e_{q_i}}\bra{e_{q_i}})$ \\
\noalign{\smallskip}
\hline
\noalign{\medskip}
\hline
\noalign{\smallskip}
\multicolumn{4}{c}{\textbf{Divergence View}} \\
\noalign{\smallskip}
\hline
\hline
\noalign{\smallskip}
VSM & $\rho_q = \proj{q}$ & $\rho_d = \proj{d}$ & $\mathcal{F}(\rho_q, \rho_d)$ \\
LM & $\rho_q = \sum_w \theta_{qw}\ket{e_w}\bra{e_w}$ & $\rho_d = \sum_w \theta_{dw}\ket{e_w}\bra{e_w}$ & $-\mathcal{VN}(\rho_q \| \rho_d)$ \\
\noalign{\smallskip}
\hline
\end{tabular*}
\end{table}

It follows from the considerations above that VSM creates query representations by accessing the whole projective space through appropriate choices of $f(w)$. On the contrary, LM ``sees'', and consequently can handle, only events from the classical sample space $\mathcal{E}$. However, the principled probabilistic foundations of the model give the flexibility of adding an arbitrary number of such events in the sequence, thus refining query representation\footnote{This is indeed the practice of Query Expansion (QE), see for example
\cite{carpineto_survey_2012}.}. In the next section, this kind of duality between VSM and LM approaches will be strengthened by analyzing the properties of the density matrices used in the two models.

Before continuing, we shall make one last consideration about the ``likelihood'' written in Eq. 1. This equation and its corresponding maximization algorithm have already been proposed by Lvovsky et al.~\cite{lvovsky_iterative_2003} in Quantum Tomography applications in order to achieve a Maximum Likelihood Estimation (MLE) of a density matrix. As we have already pointed out, $\mathcal{L}$ reduces to a classical likelihood if and only if the projectors in the sequence are picked from the same eigensystem. Therefore, the product in its general form cannot be understood as a proper likelihood. We believe that it would be interesting to focus future research in finding a proper likelihood formulation in the quantum case that would enable principled statistical estimation and Bayesian inference (see~\cite{warmuth_bayesian_2009} for a recent attempt in formulating a Bayesian calculus for density matrices).


\begin{figure}[t]
\centering
\includegraphics[scale=0.7]{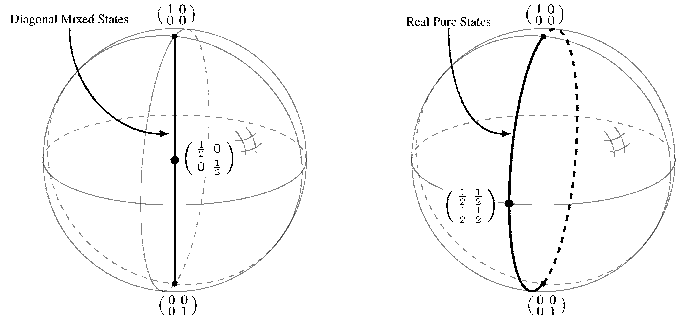}
\caption{The set $\mathcal{D}^2$ visualized using the Bloch sphere parametrization~\cite{nielsen2004}. Highlighted in black are the region of $\mathcal{D}^2$ used by LM (to the left) and VSM (to the right).}
\end{figure}

\subsection{Divergence View}
In the divergence view, a density matrix is associated both to the document and to the query and the scoring function is a divergence defined on the set $\mathcal{D}^n$ of $n \times n$ density matrices. Valuable insights can be provided by noting that the models gain access to different regions within $\mathcal{D}^n$. As an example, in Figure 1, we plot the set $\mathcal{D}^2$ using the well known Bloch parametrization~\cite{nielsen2004}. Highlighted in black are the regions of the space used by LM (to the left) and VSM (to the right). Distinct regions are likely to denote different representational capabilities.

In the case of LM, density matrices are restricted to be diagonal, i.e. mixtures over the identity eigensystem. For two density matrices to be different, one has to modify the distribution of the eigenvalues. Therefore, LM ranks based upon differences in the eigenvalues between density matrices. The picture of the VSM approach appears as the perfect dual of the preceding situation. Query and documents are represented by \emph{pure states}, i.e. dyads. Whatever the dimensionality of the Hilbert space, the mixture weights of these density matrices are concentrated onto a single projector. In order to be different, density matrices must be defined over different eigensystems. Therefore, VSM ranks based on the difference in the eigensystem between query and document density matrices.

The set of diagonal density matrices is represented in Figure 1 (left). Any two antipodal points on the surface of the sphere correspond to a particular eigensystem. Diagonal density matrices are restricted to the identity eigensystem. However, they can delve inside the sphere by spreading the probability mass across their eigenvalues. The black circle in Figure 1 (right) highlights pure states with real positive entries. These naturally lie on the surface of the Bloch sphere.

In summary, the VSM restriction to pure states leaves free choice on the eigensystem while fixing the eigenvalues. Conversely, by restricting density matrices to be diagonal, i.e. classical probability distributions, LM leaves free choice on the eigenvalues while fixing the eigensystem. Leveraging both degrees of freedom by employing the machinery of density matrices seems to be a natural step in order to achieve more precise representation for documents and queries. VSM and LM also differ in the choice of scoring functions. The former uses the Fidelity measure which is a metric on $\mathcal{D}^n$. The latter uses an asymmetric divergence on $\mathcal{D}^n$. More insights into these differences are given in the next section, where we try to contextualize our considerations by referring to common IR issues and concepts.

\section{A Joint Interpretation and Perspectives}
In~\cite{zhai_statistical_2007}, the author presents KL divergence models as ``essentially similar to the vector-space model except that text representation is based on probability distributions rather than heuristically weighted term vectors''. The analysis done in the previous section extends this remark and highlights how VSM and LM leverage very different degrees of freedom by allocating different regions in $\mathcal{D}^n$. However, no clue is given about what should be the meaning of the eigensystems and the eigenvalues from an IR point of view, nor why controlling both could be useful for IR. We will try to give some perspective for the potential usefulness of the enlarged representation space.

In basic bag-of-words retrieval models such as LM or VSM, terms are assumed to be unrelated, in the sense that each term is considered to be an atomic unit of information. To enforce this view, LM associates to each term a sample point and the VSM a dimension in a vector space. Our analysis showed that sample points correspond to dimensions in a vector space. The heritage left by LSI~\cite{deerwester_indexing_1990} suggests that a natural interpretation for such dimensions is to consider them as \emph{concepts}. In this work, we interpret projectors onto directions as concepts. Because terms are considered as unrelated, the projectors onto the standard basis $\proj{e_1}, \ldots, \proj{e_n}$ in $\mathbb{H}^n$ form a \emph{conceptual basis} in which each term labels its own underlying concept.\footnote{In~\cite{melucci_basis_2008}, \emph{each} basis of a vector space is considered as describing a \emph{contextual property} and the vectors in the basis as \emph{contextual factors}. We prefer not to adopt such interpretation for two reasons: (1) in this paper, classical sample spaces are exclusively associated to orthonormal basis and (2) we believe that referring to concepts leads to a more general formulation, better tailored to our needs.}


From this point of view, LM builds representations of queries and documents by expressing uncertainty on which concept chosen from the standard basis represents the information need. On the contrary, VSM does not have the flexibility of spreading probability weights. However, it can represent documents and queries by a unique but arbitrary concept. In VSM, the similarity score is computed by comparing how similar the query concept is to the document concept. In this picture, the cosine similarity reveals to be a measure of relatedness between concepts. In LM, the score is not at all computed on concept similarity, but by considering how the query and the document spread uncertainty on the same conceptual basis.

In order to see how this all could be instantiated, let us suppose that compound phrases such as {\em ``computer architecture''} express a different concept than \emph{``computer''} and \emph{``architecture''} taken separately. Modeling interactions between terms has been a longstanding problem in IR (for example, see~\cite{gao_dependence_2004}). We conjecture that a very natural way to handle such cases stems from our analysis. Assume that both \emph{``computer''} and \emph{``architecture''} are associated to their corresponding single term concepts, i.e. $\proj{e_c}$, $\proj{e_a}$. The concept expressed by the compound could be associated to a superposition event $\proj{k_{ca}}$ where $\ket{k_{ca}} = f(c) \ket{e_c} + f(a) \ket{e_a}$ and $f$ is a weight function (assuming normalization) expressing how compound and single term concepts are related. In this setting, the enlarged representation space turns out to be the perfect fit in order to express uncertainty on this set of concepts. One could build a density matrix associated both to a query and to a document assigning uncertainty to both single term concepts $\proj{e_c}, \proj{e_a}$ and compound concepts $\proj{k_{ca}}$. This could be done, for example, by leveraging quantum estimation methods such as described in~\cite{lvovsky_iterative_2003}. As we have pointed out before, the VN divergence could be the right scoring function in order to take into account both divergences in uncertainty distribution and concept similarities. Indeed, we have defined an IR model in this way. Details can be found in \cite{sordoni_qlm_2013}. Our experiments on several TREC collections show that the model leads to higher effectiveness than the existing models (in particular, LM).

As a last remark, we shall point out that the accounts made until now do not need the whole machinery of complex vector spaces. We do not have a practical justification for the usefulness of vector spaces defined over the complex fields (see~\cite{zuccon_use_2011} for a discussion on these issues). However, we speculate that these could bring improved representational power and thus remains an interesting direction to explore. 

\section{Conclusion}

In this work, we showed how VSM and LM can be considered dual in how they allocate the representation space of density matrices and in the nature of their scoring functions. In our interpretation, VSM adopt a symmetric scoring function which measures the concept similarity. LM fixes the standard conceptual basis and scores documents against queries based on how they spread the probability mass on such basis. We argued that leveraging both degrees of freedom could lend a more precise representations of documents and queries and could be especially effective in modelling compound concepts arising from phrasal structures. This has been confirmed by another study~\cite{sordoni_qlm_2013}. 

\bibliographystyle{plain}

\end{document}